\documentclass[a4paper,fleqn]{cas-dc}

\usepackage[numbers]{natbib}

\usepackage{gensymb}
\usepackage{tabularx}

\begin{document}
\let\WriteBookmarks\relax
\def\floatpagepagefraction{1}
\def\textpagefraction{.001}

\affiliation[LBNL]{
    organization={Lawrence Berkeley National Laboratory (LBNL)},
    city={Berkeley},
    state={California},
    country={USA},
    }

\affiliation[EPFL]{
    organization={École Polytechnique Fédérale de Lausanne (EPFL)},
    city={Lausanne},
    country={Switzerland},
    }

\affiliation[Yale]{
    organization={Yale University},
    city={New Haven},
    state={Connecticut},
    country={USA},
    }
    
\affiliation[LASP]{
    organization={Laboratory for Atmospheric and Space Physics (LASP), University of Colorado},
    city={Boulder},
    state={Colorado},
    country={USA},
    }
    
\affiliation[SSL]{
    organization={Space Sciences Laboratory (SSL), University of California Berkeley},
    city={Berkeley},
    state={California},
    country={USA},
    }

\affiliation[UW]{
    organization={University of Washington},
    city={Seattle},
    state={Washington},
    country={USA},
    }
    
\affiliation[UM]{
    organization={University of Michigan},
    city={Ann Arbor},
    state={Michigan},
    country={USA},
    }

\title[mode=title]{25,000 optical fiber positioning robots for next-generation cosmology}

\shorttitle{25,000 fiber robots for cosmology}

\shortauthors{Silber et~al.}


\author[LBNL]{Joseph H. Silber}[orcid=0000-0001-7511-2910]
\ead{jhsilber@lbl.gov}
\cormark[1] 
\cortext[cor1]{Corresponding author}

\author[LBNL]{David J. Schlegel}
\author[EPFL]{Ricardo Araujo}
\author[Yale]{Charles Baltay}
\author[LBNL]{Robert W. Besuner}
\author[LASP]{Emily Farr}
\author[LBNL]{Julien Guy}
\author[EPFL]{Jean-Paul Kneib}
\author[SSL]{Claire Poppett}
\author[UW]{Travis A. Mandeville}
\author[UM]{Michael Schubnell}
\author[EPFL]{Markus Thurneysen}
\author[UW]{Sarah Tuttle}

\begin{abstract}
    Massively parallel multi-object spectrographs are on the leading edge of cosmology instrumentation. The highly successful Dark Energy Spectroscopic Instrument (DESI) which begun survey operations in May 2021, for example, has 5,000 robotically-actuated multimode fibers, which deliver light from thousands of individual galaxies and quasars simultaneously to an array of high-resolution spectrographs off-telescope. The redshifts are individually measured, thus providing 3D maps of the Universe in unprecedented detail, and enabling precise measurement of dark energy expansion and other key cosmological parameters. Here we present new work in the design and prototyping of the next generation of fiber-positioning robots. At 6.2\,mm center-to-center pitch, with 1-2\,$\mu$m positioning precision, and in a scalable form factor, these devices will enable the next generation of cosmology instruments, scaling up to instruments with 10,000 to 25,000 fiber robots.
\end{abstract}

\begin{keywords}
    fiber positioner robot\sep
    optical fiber focal plane\sep
    multi-object spectrograph \sep
    dark energy cosmology
\end{keywords}

\maketitle

\section{Introduction}
    
    Three-dimensional maps of the locations of galaxies in the Universe have provided a powerful approach to measuring the clustering of galaxies, which in turn measures the effects of dark matter, dark energy and gravity.  The earliest such map was the Center for Astrophysics Redshift Survey, which mapped the redshift-space positions of 2,401 galaxies.\citep{Huchra1983} This was painstaking work, observing one galaxy at a time between 1977 and 1982.
    
    The advent of multi-fiber spectroscopy has revolutionized our ability to map large volumes of the Universe. Telescopes focus light onto a virtual focal plane, which in general is a curved surface. Instrumentation was designed to simultaneously capture light from many galaxies on that surface by placing the tip of a multi-mode optical fiber at the location of each galaxy on that surface to collect its light. The first large-scale implementation of this technique was with the Two-degree-Field Galaxy Redshift Survey (2dF) on the 3.9m Anglo-Australian Telescope \citep{Colless2001} and the Sloan Digital Sky Survey (SDSS; \citep{Smee13}) on the 2.5m Sloan Telescope. Those telescopes fielded instruments that could position 392 and 640 fibers, respectively, on their focal planes to map up to that many galaxies in each observation. More recently, the Dark Energy Spectroscopic Instrument (DESI) was commissioned on the 4m Mayall Telescope in 2019-2020 and represents the most powerful such instrument, mapping up to 5,000 galaxies in each exposure.\citep{DESIOverviewPreprint}
    
\begin{figure}
    	\centering
    		\includegraphics[width=\columnwidth]{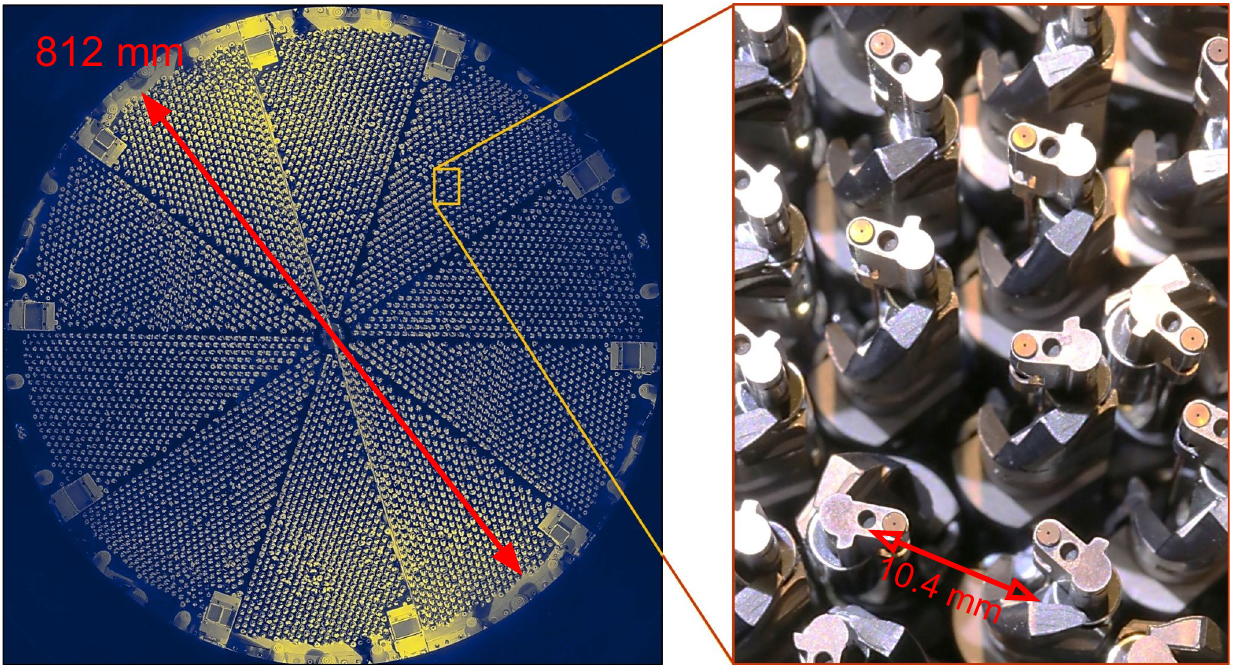}
    	\caption{\emph{Left:} Photograph through the optical corrector of the current state-of-the-art DESI focal plane. It has 5,000 fiber robots, mounted on 10 independent 36\degree~modules. \emph{Right:} close-up view of the DESI robots, which have a minimum 10.4 mm mounting pitch.}
    	\label{fig:desi_focal_plane}
    \end{figure}

    A critical component to these multi-fiber instruments has been the mechanism for rapidly re-positioning the location of the fibers on the focal plane to the position of the galaxies in each telescope pointing.  Earlier instruments made use of manually re-positioned fibers in plug plates (SDSS) or robotic pick-and-play systems that would reposition fibers serially (2dF).  DESI is a robotic system (Figure~\ref{fig:desi_focal_plane}) that re-configures all 5,000 fibers simultaneously in less than two minutes.\citep{DESIFocalPlanePreprint} The work presented in this paper will enable a scaling of the DESI approach to 10,000 fibers on the DESI focal plane, or up to 25,000 fibers on the MegaMapper focal plane.\citep{Schlegel2019}
    
    DESI has demonstrated the performance of a robotic system with individually-actuated robots re-positioning each fiber prior to an exposure. In its first year of operation, DESI has measured redshift distances to 14 million galaxies, easily superseding the sum of all previous galaxy maps.  This paper presents a further miniaturization of fiber positioning robots, reducing the center-to-center pitch from 10.4\,mm to 6.2\,mm, while maintaining the key performance characteristics of speed and positioning accuracy.  A system with these smaller robots on the DESI focal plane would double the number of fibers from 5,000 to 10,000, essentially doubling the performance of that instrument.
    
\begin{figure}
    	\centering
    		\includegraphics[width=\columnwidth]{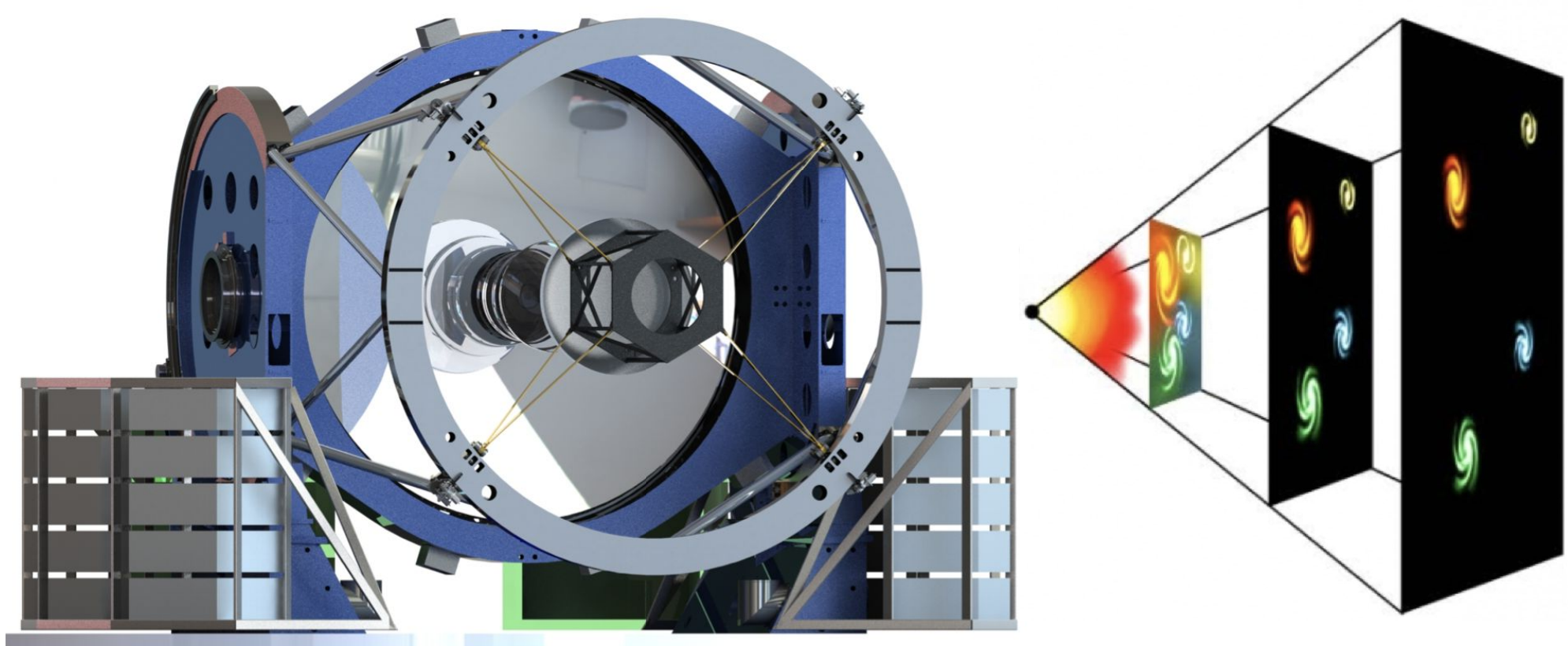}
    	\caption{Several telescopes are being analyzed to receive a multi-object spectrograph instrument as discussed in this paper. Significant interest is building in the community to build ``MegaMapper'', a dedicated telescope with a 6.5\,m primary mirror. [Image credit: Jeff Crane (Carnegie Observatories)]}
    	\label{fig:megamapper_telescope}
    \end{figure}

    The MegaMapper (Figure~\ref{fig:megamapper_telescope}) telescope's focal plane is physically larger and could accommodate 25,000 of these robots.  In combination with a larger primary mirror, MegaMapper would represent a 15-fold increase in performance compared to DESI.

\section{System architecture}

\begin{figure}
    	\centering
    		\includegraphics[width=\columnwidth]{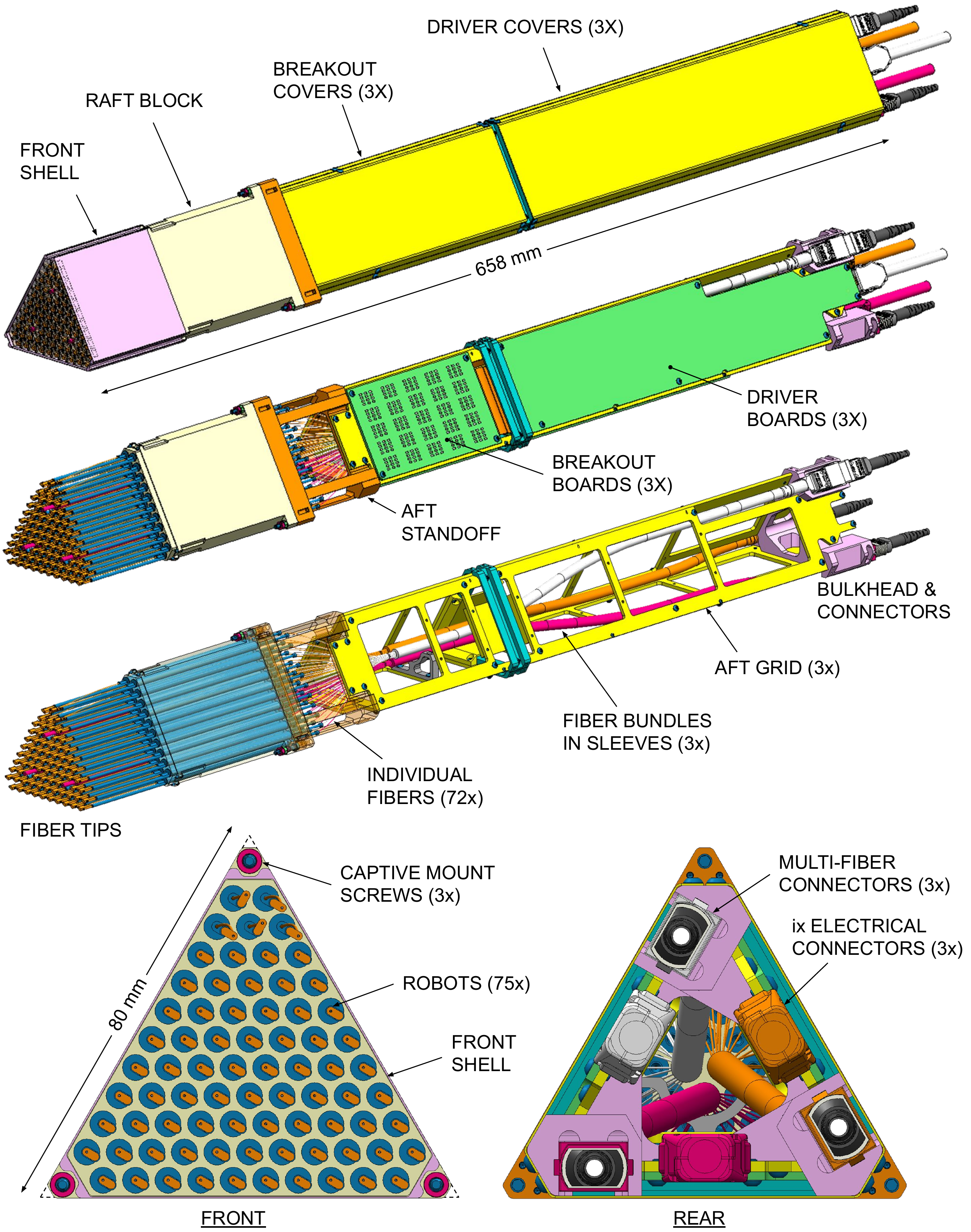}
    	\caption{Each raft module is a complete miniature instrument, including 75 robotically-positioned fibers and electronics, and weighing $\sim$\,1.7\,kg. Internally the raft is further subdivided into 3 logical and mechanical groups of 25 fibers / robots each. At the rear bulkhead are electrical and fiber connectors. The module mounts to the focal plate by 3 captive M3 screws at the corners, including captive shims for precise alignment of focus, tip, and tilt.}
    	\label{fig:raft_overview}
    \end{figure}

    We are designing a modular fiber positioning system, in which each ``raft'' module (Figure~\ref{fig:raft_overview}) can target 75 locations with 75 fibers simultaneously. Each robotic fiber unit within the raft is an individual SCARA-like mechanism, with two rotation axes, for a total of 150 degrees of freedom.
    
    General behaviour and performance will be comparable to robots built for the DESI and SDSS-V projects. However, at 6.2\,mm center-to-center pitch, the fiber density will be much higher than either of these predecessors (DESI\,=\,10.4\,mm, SDSS-V\,=\,22.4\,mm).
    
    Previous instruments of this type have generally been composed of independent actuator units, with one fiber per unit. Hundreds of these units are then integrated into a single mounting plate, to form the focal plane. In our experience, the DESI and SDSS-V projects (500 such positioning units per mounting plate, with DESI having ten such plates and SDSS-V having one) have been approaching the practical limits of instrument complexity. Full-up testing takes years to achieve and maintenance scenarios become major refurbishment efforts.
    
    For the next generation of instrument, we intend that each of our 75-fiber raft modules should be a complete working miniature instrument, including robotics, fibers, support structure, and electronics. One will be able to plug it in on one’s desktop and fully operate the robots. Installation and extraction from the larger mounting plate at the telescope will be rapid ($\sim$\,15\,min per unit), and will not require the heavy support equipment of earlier generations.\citep{Besuner2020} On-site spares will be relatively inexpensive, shipment of raft modules will be low cost and low risk, and rework will be practical for a single technician to do on a lab bench.
    
\begin{figure}
    	\centering
    		\includegraphics[width=\columnwidth]{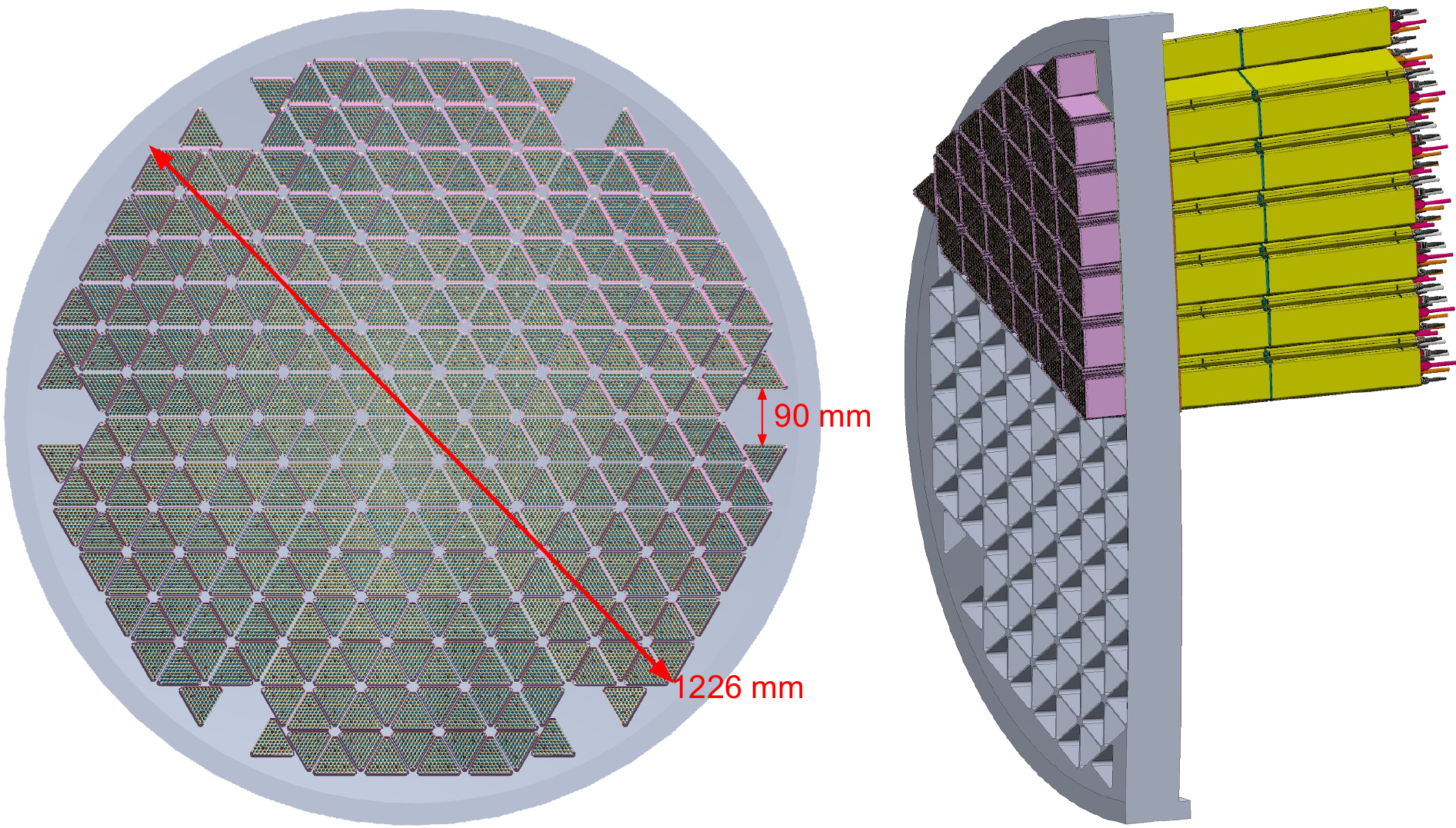}
    	\caption{\emph{Left:} 3D CAD model of the MegaMapper focal plate layout. Six open pockets at the perimeter provide efficient locations for guide cameras up to 90 mm wide. The outer radius of the physical plate will likely be on the order of R638--675 mm or so. \emph{Right:} 58 rafts are shown populating one 60\degree~region. }
    	\label{fig:megamapper_layout}
    \end{figure}

    We ultimately intend to build focal plane systems composed of $\sim$100--350 rafts (quantity depending on each particular telescope's size), with dozens of spare rafts on hand. The focal planes will generally consist of a metal grid into which the rafts are mounted very close to each other, at an expected spacing of 1 to 3 mm. A layout of 348 rafts for the proposed MegaMapper telescope is shown in Figure~\ref{fig:megamapper_layout}.
    
    Several architectural questions emerge from decisions regarding connecting the short fiber length in the raft to the longer fiber cable run to the spectrographs. We are currently considering three main options.
    
    First, the most logistically convenient method would be to have multi-fiber connectors at the rear bulkhead of each raft. We anticipate that the 75 fibers would be arranged in 3 harnesses. Each harness would have a single connector loaded with 25 fibers. These would be mounted at a bulkhead at the rear of the raft. We have designed the raft to ensure space for this, however connectors can cost significant throughput. We are actively evaluating whether a sufficiently performant connector solution is possible.
    
    Second, as was done on the successful DESI instrument, we could fusion splice fibers at the focal plane to the larger fiber cables. This is a slower and less convenient option for raft replacement. The fused fibers have excellent throughput performance. On DESI, the fiber system had $\ge$~90\% total throughput from the prime focus corrector to the spectrograph \citep{poppett20}, a very high number.
    
    Finally, we are investigating the possibility of a single fiber cable per raft, all the way to the spectrograph. With no connector or splice, this would be the most performant option, but its feasibility depends especially on the particular telescope size and kinematics. We are actively working on reduced-size fiber slit heads that would enable plugging rafts independently into DESI-style spectrographs.

\section{Fiber positioning}

    The raft architecture could accommodate various fiber actuation implementations. We are currently basing our design on 2 DOF SCARA-like mechanisms, similar to what has been successfully deployed on DESI and SDSS-V. The central rotation axis of each robot is called $\alpha$ (alpha) and the eccentric axis is called $\beta$ (beta). As on DESI, each axis will be driven by a \o4\,mm DC brushless gearmotor.
    
\begin{figure}
    	\centering
    		\includegraphics[width=\columnwidth]{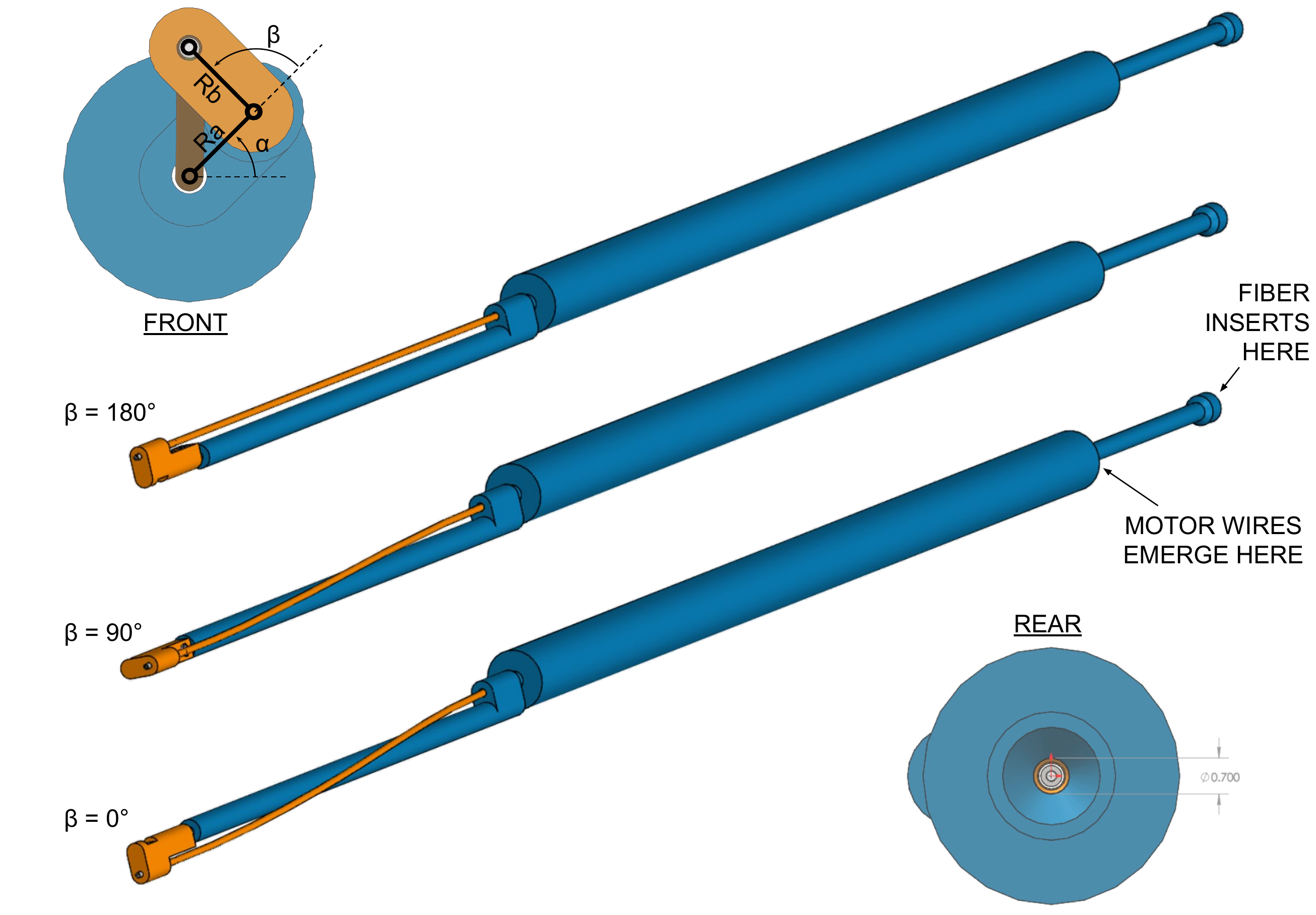}
    	\caption{A cartoon of the individual fiber robot (intentionally not showing particularities of internal mechanics) shows the kinematics, spacing, and approximate size and shape, around which the Raft is being designed.}
    	\label{fig:dummy_robot_illustration}
    \end{figure}

    A cartoon showing the general kinematics, fiber route, and aspect ratios is given in Figure~\ref{fig:dummy_robot_illustration}. Both in order to achieve tight center-to-center pitch (i.e. between adjacent robots' $\alpha$ axes), and to ensure gentle fiber routes (as on DESI, we adopt 50\,mm as our minimum fiber bend radius throughout) these devices tend to be long and narrow.
    
    The motors do not require encoders. Better positioning is obtained via a Fiber View Camera (FVC) system.\citep{baltay19} This is a high-resolution camera with a telescopic lens, pointing through the optical corrector at the focal plane. During reconfiguration of the fibers, we back-illuminate the fibers with bright LEDs at the spectrographs. The fiber positions can be centroided to 30 millipixel precision. On DESI, we achieved 3\,$\mu$m\,RMS precision with such a system. First we perform a ``blind'' move (for which DESI has $54 \mu$m\,RMS error) to get close to the target position. The true fiber positions are measured with the FVC and a single set of correction moves is executed, bringing the thousands of fibers to within $6 \mu$m\,RMS of their commanded targets. The motors are sufficiently geared down that the FVC measurement plus variable air turbulence effects in the telescope dome are in fact the limiting factors for positioning precision.
    
    Positioning accuracy is dependent upon our ability to map focal plane coordinates to astrometric sky coordinates. A combination of guide cameras (to view the sky) and fixed fiducial point sources (to locate the guide cameras in the FVC image) makes this linkage. DESI has 2\,$\mu$m accuracy in low-turbulence conditions with this type of system, and in general is well under 10\,$\mu$m in all conditions. We look forward to the new engineering challenges that surely will result when scaling up from 5,000 to 25,000 fibers per readout, but there are no real show stoppers.
    
    The robots have the ability to reach within their neighbors' patrol envelopes. Collisions between robots do not cause damage, but do prevent those fibers from achieving their targets. To prevent such collisions, we pre-calculate non-colliding, timed schedules for the rotations of all motors. These schedules are pushed to the respective driver electronics, which are then commanded to start executing their moves upon a synchronized start signal. 

\section{Fiber handling and alignment}
    
    At the front of the raft, each fiber will be individually retained by the fiber arm of a given robot. The mounting of fibers is technically challenging, because even what intuitively seem like gentle mechanical stresses can significantly degrade the fiber’s optical performance, by inducing Focal Ratio Degradation (FRD). We also must control the fiber angle with respect to the optical corrector's local chief ray to within fractions of a degree.
    
    Our experience with attaching fibers to the robots in previous experiments has never been entirely satisfactory from a maintenance and logistics standpoint. For example in DESI, we robotically focused each fiber individually in its robot, and then glued the fiber in position. Bonding the fiber directly to the robot imposes a significant logistical constraint, and can cost valuable hardware when any one part fails. We are currently doing R\&D on better methods to hold fibers in these tiny actuators (whose end effectors are 0.8\,mm radius), for example with magnetically attracted ferrules and built-in auto-focusing mechanims. These represent interesting design,  engineering, and manufacturing challenges which we hope to solve in the next couple of years.
    
    A key requirement we are imposing is that fibers be smoothly and easily insertable from the back side of the robots. We anticipate having a flexible plastic tube running through the length of the robot from the rear to the $\beta$ arm, through which the fiber can be inserted blindly all the way through the mechanism.

\section{Electronics}
    We will divide the raft electronics generally into two types of boards: ``breakout'' versus ``driver''. There are three pairs of such boards per raft, again following the architectural design rule of dividing the equilateral triangle-shaped raft into 3 logical groups of 25 robots each. The 50 motors in each group will be connected to their respective breakout board. The breakout board will be as simple as possible, with no active components. The active electronics will all be on the driver board, which will plug into the breakout board via a single board-to-board connector.
    
    At the rear end of each driver board will be a standard ix Industrial Connector, providing power and Ethernet (Type 4 PoE++, up to 71\,W delivered power) for those 25 robots.

\begin{figure}
    	\centering
    		\includegraphics[width=\columnwidth]{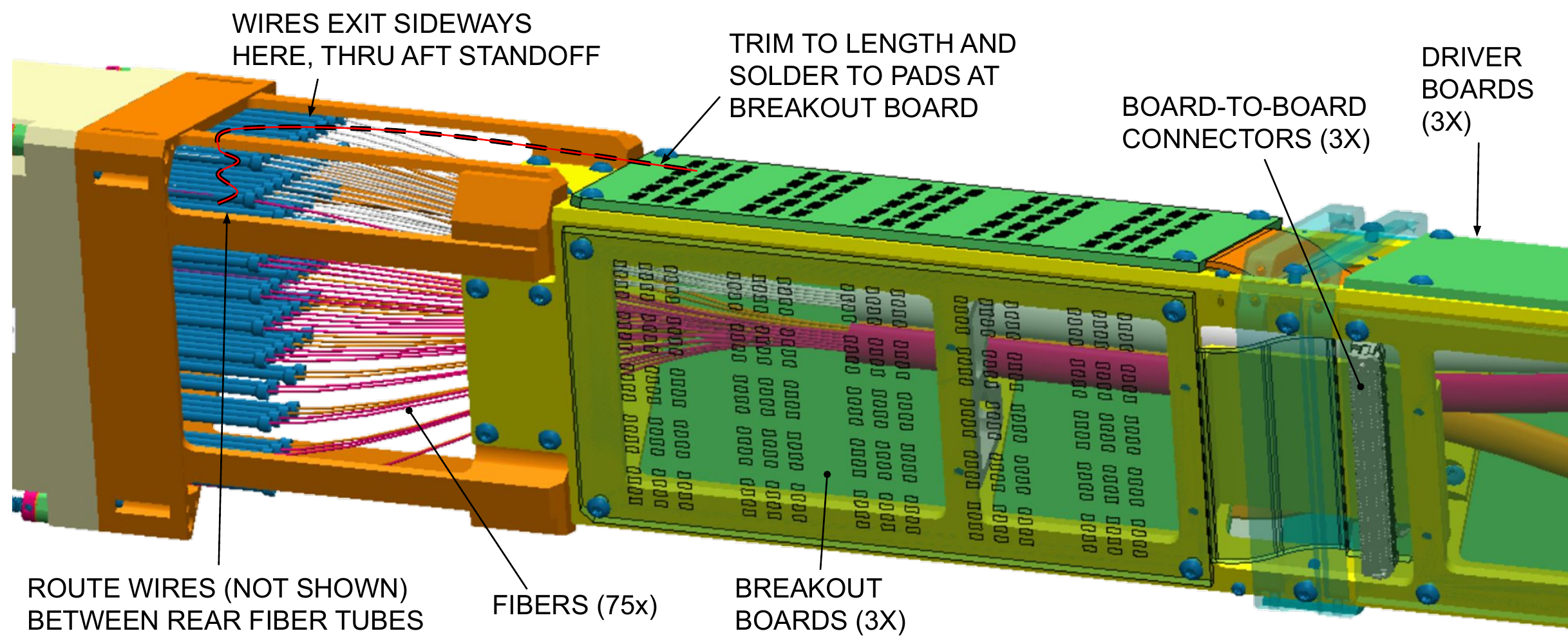}
    	\caption{Assembly concept for the most complex zone of the raft module, where 75 fibers must bypass 450 motor wires. (Sensorless motors have 3 leads per motor x 2 motors per robot. Use of sensors would increase this wire count by 4 leads per robot or more, depending on design details.) The raft design allows independent insertion / removal of fibers, accessible test points for all motor leads, and easy replacement of driver electronics without disturbing motor wire connections.}
    	\label{fig:raft_electronics_and_fiber_weave}
    \end{figure}

    At the rear of the robots, the raft will have a dense array of 75 individual fibers and hundreds of wires which must be routed through each other. Our assembly concept, illustrated in Figure~\ref{fig:raft_electronics_and_fiber_weave}, is that the motor wires will emerge from the robot forward of the fiber insertion tube. The wires will wrap through the array of tubes, emerging out the side, where they go up to the breakout board. This design ensures that fiber installation / removal can be independent of electrical wires. Note that in the dummy model depicted in the Reference Design, this rear fiber tube is shown centered, but can be shifted to the side as necessary for a given robot design.

\section{Prototypes}

\begin{figure}
    	\centering
    		\includegraphics[width=\columnwidth]{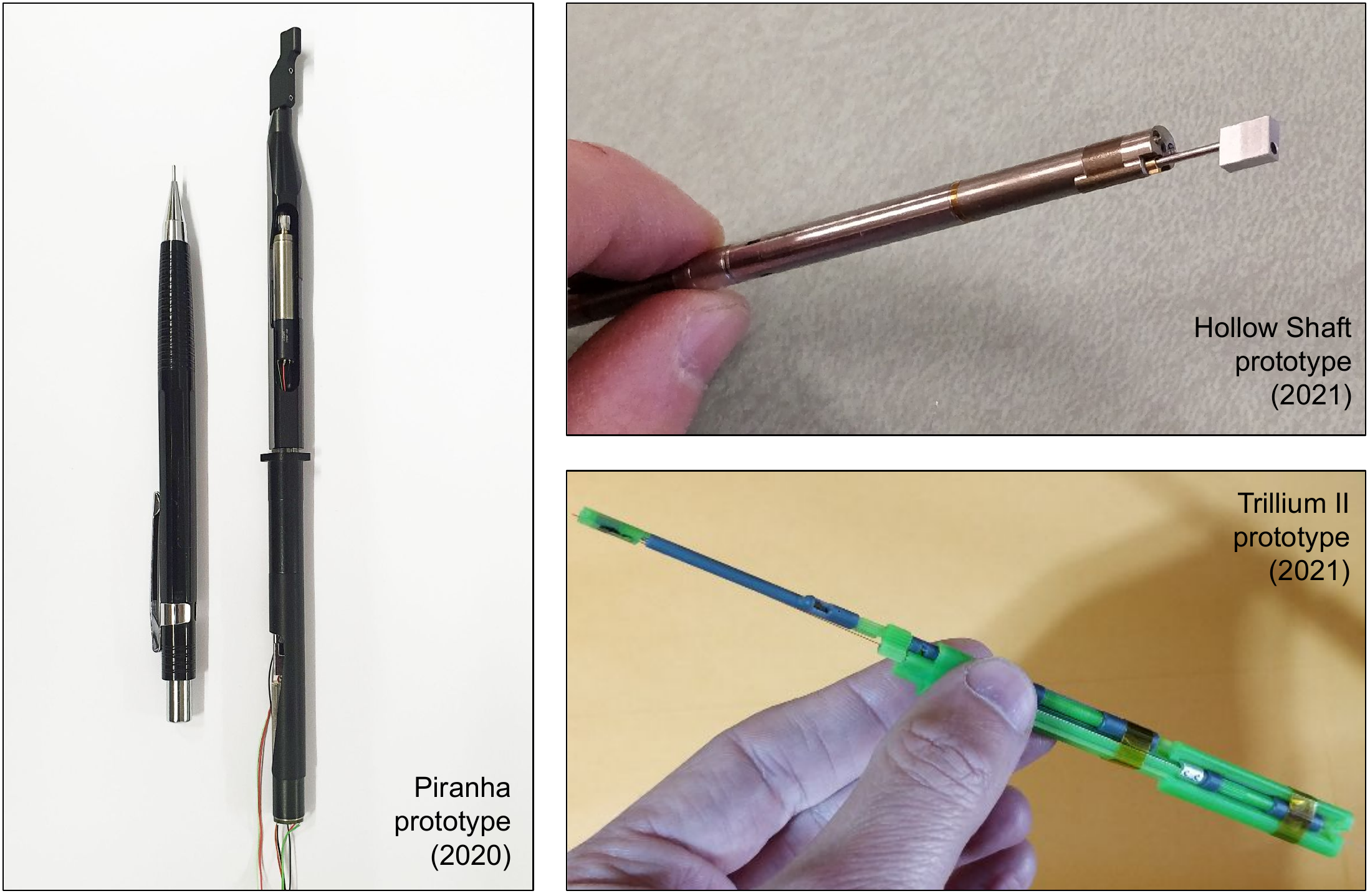}
    	\caption{Recent developments in further miniaturization of DC brushless based fiber positioner robots. The \o9.5\,mm Piranha design (left) was developed by \'Ecole Polytechnique F\'ed\'erale de Lausanne (EPFL) together with the Swiss manufacturer of miniature microsystems, MPS Microsystems, with 8 units produced \citep{kronig2020}. Relatively large motors were used in these prototypes but these can be replaced to scale down the mounting pitch of the design. The Hollow-Shaft units (upper right) were designed and built by Adamant Namiki Precision Jewel Co., to meet an open specification and RFQ from a consortium of investigators at EPFL, Lawrence Berkeley National Lab (LBNL), and the University of Michigan (UM). The hollow-shaft design is \o5\,mm plus an additional 0.5\,mm radial bulge near the tip. Finally, several examples of a triple-fiber design (Trillium, lower right) were built and tested at LBNL. These incorporate six off-the-shelf \o4\,mm DC brushless motors into a minimal set of plastic parts (presumably lending themselves ultimately to mass-manufacturing via injection-molding), and have MegaMapper's target 6.2 mm spacing.}
    	\label{fig:piranha_hollowshaft_trillium}
    \end{figure}

    Our collaboration has produced several important prototypes in the last two years. Examples are shown in Figure~\ref{fig:piranha_hollowshaft_trillium}.
    
    Accuracy of the prototypes was encouraging. `Piranha' prototypes positioned within $\le 5\,\mu$m over their full range after two correction moves, `Trillium' had 1\,$\mu$m repeatability over most of its range, with fixable issues near hard limit extrema, and the `Hollow-Shaft' design had 7--12\,$\mu$m\,RMS repeatability, a commendable result considering that, to our knowledge, no such hollow-shafted DC brushless gearmotor at this size scale had previously been built. We thus have confidence in the basic principle of the raft's spacing and actuation.

\begin{figure}
    	\centering
    		\includegraphics[width=\columnwidth]{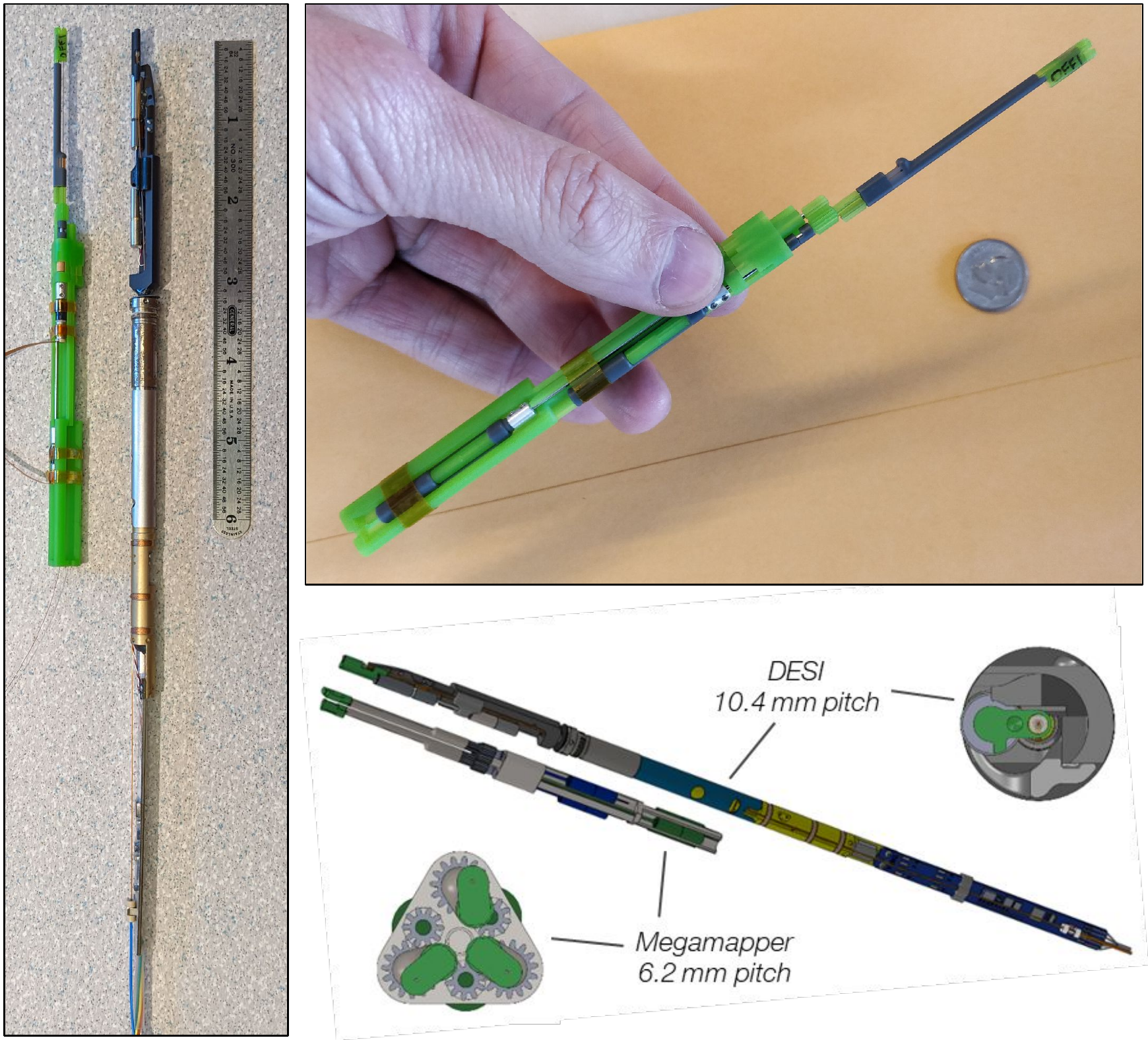}
    	\caption{\emph{Left and bottom-right:} A Trillium fiber positioning robot is shown in comparison to the DESI model. \emph{Upper-right:} A Trillium prototype is shown in comparison to a U.S. dime (\o 17.9 mm). Note: the Trillium prototypes shown in these views have only one of the three fiber actuator positions populated. This simplifies basic performance testing by not requiring anticollision move scheduling among the three neighbors.}
    	\label{fig:trillium_size}
    \end{figure}

    The Trillium design (Figure~\ref{fig:trillium_size}) was developed in particular to test key assumptions for the larger raft design. We built two generations of Trillium prototypes. The design integrates 6 motors, driving 3 fibers, into a single unit. Our goals were to explore (1) the tightest possible center-to-center pitch that we could achieve with conventional \o4\,mm motors, (2) minimize part count, (3) eliminate glue joints, and (4) maximize the potential for injection molding of components.

    In addition to the size reduction (Trillium demonstrates 2.8x higher fiber density over the current state-of-the-art DESI instrument), the prototypes demonstrated configurations with two advantages over previous designs: (1) both gearmotors and their wires are fixed in place, rather than rotating with the fiber arm, and (2) the fibers route easily and safely through a central hole.

\section{Specification and 3D models}

    In spring 2022, our group released an open specification for the raft, accompanied by a detailed Reference Design (RD) concept as a 3D CAD assembly model. An abbreviated list of several key specifications is shown in Table \ref{tab:specs}.
    
    The specification document may be downloaded at \url{https://zenodo.org/record/6986668} and the model at \url{https://zenodo.org/record/6354853}. The top-level assembly is called “MM Raft Assembly”. Several vendors have responded with enthusiasm, and we look forward to testing their implementations.
    
    We additionally provided a 3D CAD model of the Trillium prototype design alongside the specification, which may be downloaded at \url{https://zenodo.org/record/6354859}. The top-level assembly is called “Trillium2 Robot”. All tooling fixture designs are included in the Solidworks version of the supplied CAD model.
    
    \begin{table*}
        \caption{Several key specifications for Raft modules.}\label{tab:specs}
        \begin{tabular}{l l}
        \toprule
        Title & Requirement or feature \\
        \midrule
        Envelope & Equilateral triangle with 80\,mm sides, chamfered\,by 2.5 mm at corners \\
        Length \& mass & $\le$~800\,mm, $\le$~2\,kg \\
        Robot spacing & 6.2\,mm center-to-center \\
        Kinematic arm lengths & $R_\alpha = 1.8$\,mm (from central axis to eccentric axis) \\
        ~ & $R_\beta = 1.8$\,mm (from eccentric axis to fiber center) \\
        Ranges of motion & -5~$\le \alpha \le$~365\degree \\
        ~ & -5~$\le \beta \le$~185\degree \\
        Hard stops & both axes \\
        Min clearance for fiber pass-through & \o0.7\,mm \\
        Min fiber bend radius & $\le$\,50\,mm \\
        Power consumption & $\le$\,1.2\,W per motor at full speed \\
        XY positioning precision & $\le$\,5\,$\mu$m\,RMS \\
        XY positioning accuracy & $\le$\,50\,$\mu$m\,RMS \\
        Fiber defocus & $\le 50 \mu$m, for all $(\alpha, \beta)$ \\
        Fiber tilt & $\le 0.5$\degree, for all $(\alpha, \beta)$ \\
        Speed at output shaft & 180\degree/sec (goal), 30\degree/sec (minimum) \\
        Lifetime & $\ge 100,000$ move cycles \\
        Operating conditions & Temperature: -20\degree C to +40\degree C, Humidity: $\sim$ 0 to 80\% RH \\
        Survival conditions & Temperature: -30\degree C to +60\degree C, Humidity: $\sim$ 0 to 95\% RH \\
        \bottomrule
        \end{tabular}
    \end{table*}
    
\section{Conclusion}

    The next generation of cosmology surveys using multi-object fiber-fed spectrographs will require a major scale-up from the current, already massively parallelized, state-of-the-art. Based on our experience constructing, commissioning, and operating the DESI and SDSS-V instruments, we have developed a modularized architecture which we believe will meet this challenge. Fiber robots capable of 1--2\,$\mu$m~positioning will be integrated in robust, triangular-shaped `raft' modules, suitable for mass-production and serviceability. We see clear paths to 10,000-fiber and 25,000-fiber systems, (1) using the existing DESI corrector at the Mayall Telescope, located atop Iolkam Du’ag (Kitt Peak) in the Tohono O’odham Nation (near Tucson, Arizona), and (2) building a new MegaMapper telescope in the Southern Hemisphere, respectively. We look forward to these exciting future projects and welcome new collaborating scientists, technicians, and engineers to make this a reality.

\section*{Acknowledgments}
    This research is supported by the Director, Office of Science, Office of High Energy Physics of the U.S. Department of Energy under Contract No. DE–AC02–05CH11231, and by the universities and institutions listed for the collaborating authors.

\bibliographystyle{cas-model2-names}
\bibliography{cas-refs}{}

\begin{thebibliography}{10}
\expandafter\ifx\csname natexlab\endcsname\relax\def\natexlab#1{#1}\fi
\providecommand{\url}[1]{\texttt{#1}}
\providecommand{\href}[2]{#2}
\providecommand{\path}[1]{#1}
\providecommand{\DOIprefix}{doi:}
\providecommand{\ArXivprefix}{arXiv:}
\providecommand{\URLprefix}{URL: }
\providecommand{\Pubmedprefix}{pmid:}
\providecommand{\doi}[1]{\href{http://dx.doi.org/#1}{\path{#1}}}
\providecommand{\Pubmed}[1]{\href{pmid:#1}{\path{#1}}}
\providecommand{\bibinfo}[2]{#2}
\ifx\xfnm\relax \def\xfnm[#1]{\unskip,\space#1}\fi
\bibitem[{Abareshi et~al.(2022)}]{DESIOverviewPreprint}
\bibinfo{author}{Abareshi, B.}, et~al., \bibinfo{year}{2022}.
\newblock \bibinfo{title}{Overview of the instrumentation for the dark energy
  spectroscopic instrument}.
\newblock \URLprefix \url{https://arxiv.org/abs/2205.10939},
  \DOIprefix\doi{10.48550/ARXIV.2205.10939}.
\bibitem[{{Baltay} et~al.(2019){Baltay}, {Rabinowitz}, {Besuner}
  et~al.}]{baltay19}
\bibinfo{author}{{Baltay}, C.}, \bibinfo{author}{{Rabinowitz}, D.},
  \bibinfo{author}{{Besuner}, R.}, et~al., \bibinfo{year}{2019}.
\newblock \bibinfo{title}{{The DESI Fiber View Camera System}}.
\newblock \bibinfo{journal}{Publications of the Astronomical Society of the
  Pacific} \bibinfo{volume}{131}, \bibinfo{pages}{065001}.
\newblock \DOIprefix\doi{10.1088/1538-3873/ab15c2}.
\bibitem[{Besuner et~al.(2020)}]{Besuner2020}
\bibinfo{author}{Besuner, R.}, et~al., \bibinfo{year}{2020}.
\newblock \bibinfo{title}{{Installation of the Dark Energy Spectroscopic
  Instrument at the Mayall 4-meter telescope}}, in: \bibinfo{editor}{Evans,
  C.J.}, \bibinfo{editor}{Bryant, J.J.}, \bibinfo{editor}{Motohara, K.} (Eds.),
  \bibinfo{booktitle}{Ground-based and Airborne Instrumentation for Astronomy
  VIII}, \bibinfo{organization}{International Society for Optics and
  Photonics}. \bibinfo{publisher}{SPIE}. p. \bibinfo{pages}{1144710}.
\newblock \URLprefix \url{https://doi.org/10.1117/12.2561507},
  \DOIprefix\doi{10.1117/12.2561507}.
\bibitem[{Colless et~al.(2001)}]{Colless2001}
\bibinfo{author}{Colless, M.}, et~al., \bibinfo{year}{2001}.
\newblock \bibinfo{title}{{The 2dF Galaxy Redshift Survey: spectra and
  redshifts}}.
\newblock \bibinfo{journal}{Monthly Notices of the Royal Astronomical Society}
  \bibinfo{volume}{328}, \bibinfo{pages}{1039--1063}.
\newblock \DOIprefix\doi{10.1046/j.1365-8711.2001.04902.x},
  \href{http://arxiv.org/abs/astro-ph/0106498}{\tt arXiv:astro-ph/0106498}.
\bibitem[{{Huchra} et~al.(1983){Huchra}, {Davis}, {Latham} and
  {Tonry}}]{Huchra1983}
\bibinfo{author}{{Huchra}, J.}, \bibinfo{author}{{Davis}, M.},
  \bibinfo{author}{{Latham}, D.}, \bibinfo{author}{{Tonry}, J.},
  \bibinfo{year}{1983}.
\newblock \bibinfo{title}{{A survey of galaxy redshifts. IV - The data}}.
\newblock \bibinfo{journal}{Astrophysical Journal, Supplement}
  \bibinfo{volume}{52}, \bibinfo{pages}{89--119}.
\newblock \DOIprefix\doi{10.1086/190860}.
\bibitem[{Kronig et~al.(2020)Kronig, Caseiro, Hug, Charif, Bouri and
  Kneib}]{kronig2020}
\bibinfo{author}{Kronig, L.}, \bibinfo{author}{Caseiro, S.},
  \bibinfo{author}{Hug, M.}, \bibinfo{author}{Charif, M.},
  \bibinfo{author}{Bouri, M.}, \bibinfo{author}{Kneib, J.P.},
  \bibinfo{year}{2020}.
\newblock \bibinfo{title}{{An easily scalable Theta/Phi fiber positioner to
  reduce risks, lead times, and costs for multi-object spectrographs}}, in:
  \bibinfo{editor}{Navarro, R.}, \bibinfo{editor}{Geyl, R.} (Eds.),
  \bibinfo{booktitle}{Advances in Optical and Mechanical Technologies for
  Telescopes and Instrumentation IV}, \bibinfo{organization}{International
  Society for Optics and Photonics}. \bibinfo{publisher}{SPIE}. pp.
  \bibinfo{pages}{435 -- 443}.
\newblock \URLprefix \url{https://doi.org/10.1117/12.2562717},
  \DOIprefix\doi{10.1117/12.2562717}.
\bibitem[{Poppett et~al.(2020)}]{poppett20}
\bibinfo{author}{Poppett, C.}, et~al., \bibinfo{year}{2020}.
\newblock \bibinfo{title}{{Performance of the Dark Energy Spectroscopic
  Instrument (DESI) fiber system}}, in: \bibinfo{booktitle}{Society of
  Photo-Optical Instrumentation Engineers (SPIE) Conference Series}, p.
  \bibinfo{pages}{1144711}.
\newblock \DOIprefix\doi{10.1117/12.2562565},
  \href{http://arxiv.org/abs/2101.11564}{\tt arXiv:2101.11564}.
\bibitem[{{Schlegel} et~al.(2019){Schlegel}, {Kollmeier} and
  {Ferraro}}]{Schlegel2019}
\bibinfo{author}{{Schlegel}, D.}, \bibinfo{author}{{Kollmeier}, J.A.},
  \bibinfo{author}{{Ferraro}, S.}, \bibinfo{year}{2019}.
\newblock \bibinfo{title}{{The MegaMapper: a z>2 spectroscopic instrument for
  the study of Inflation and Dark Energy}}, in: \bibinfo{booktitle}{Bulletin of
  the American Astronomical Society}, p. \bibinfo{pages}{229}.
\newblock \href{http://arxiv.org/abs/1907.11171}{\tt arXiv:1907.11171}.
\bibitem[{Silber et~al.(2022)Silber, Fagrelius, Fanning
  et~al.}]{DESIFocalPlanePreprint}
\bibinfo{author}{Silber, J.H.}, \bibinfo{author}{Fagrelius, P.},
  \bibinfo{author}{Fanning, K.}, et~al., \bibinfo{year}{2022}.
\newblock \bibinfo{title}{The robotic multi-object focal plane system of the
  dark energy spectroscopic instrument (desi)}.
\newblock \URLprefix \url{https://arxiv.org/abs/2205.09014},
  \DOIprefix\doi{10.48550/ARXIV.2205.09014}.
\bibitem[{Smee et~al.(2013)}]{Smee13}
\bibinfo{author}{Smee, S.A.}, et~al., \bibinfo{year}{2013}.
\newblock \bibinfo{title}{The multi-object, fiber-fed spectrographs for the
  sloan digital sky survey and the baryon oscillation spectroscopic survey}.
\newblock \bibinfo{journal}{Astronomical Journal} \bibinfo{volume}{146},
  \bibinfo{pages}{32}.
\newblock \DOIprefix\doi{10.1088/0004-6256/146/2/32},
  \href{http://arxiv.org/abs/1208.2233}{\tt arXiv:1208.2233}.

\end{thebibliography}

\end{document}